\documentclass[aps,prl,twocolumn,groupedaddress,showpacs]{revtex4-1}
\usepackage{bm}
\usepackage{graphicx}
\usepackage{epstopdf}
\usepackage{amsmath}

\begin{document}


\title{The centrifugal Jahn-Teller effect in rare-earth cage systems }


\author{M. Amara}
\email{mehdi.amara@neel.cnrs.fr}
\affiliation{Univ. Grenoble Alpes, Inst NEEL, F-38000 Grenoble, France}
\affiliation{CNRS, Inst NEEL, F-38000 Grenoble, France}


\date{\today}

\begin{abstract}
The analysis of the properties of rare-earth based materials relies on the Crystalline Electric Field theory. This theory has to be reconsidered in case of cage-type compounds, where the rare-earth ion can substantially depart from its, high symmetry, average position. It is shown that, in case of an orbital degeneracy at the cage center, a specific Jahn-Teller effect develops in the paramagnetic range: at low temperature, the distribution of the magnetic ion spreads inside the cage, the magnetic entropy is reduced, whereas the cage multipolar susceptibilities are increased.  These consequences are put in relation with the properties of some metallic rare-earth cage compounds.
\end{abstract}


\maketitle

\section{}

Over the last decade, impulsed by an interest in thermoelectric applications, the investigation of metallic cage compounds has soared. In these systems, atoms are enclosed in oversized cages, allowing large excursions from their average positions. In some crystallographic structures, the cage can accommodate a rare-earth ion, giving rise to specific magnetic properties. The most investigated rare-earth cage compounds are filled skutterudites, that crystallize according to the LaFe$_4$P$_{12}$-type structure \cite{Jeitschko1977}. These compounds display a variety of intriguing features, as the heavy fermion and superconductor PrOs$_4$Sb$_{12}$ \cite{Bauer2002}, or the metal-insulator transition (MI) in PrRu$_4$P$_{12}$ \cite{Iwasa2005}, the non-magnetic ordering of PrFe$_4$P$_{12}$ \cite{Keller2001}, etc.. Recently, many other promising rare-earth cage compounds have emerged under the CeCr$_2$Al$_{20}$-type structure \cite{Niemann1995, Burnett2014}, notably praseodymium compounds that order non-magnetically \cite{Onimaru2011, Sato2012}. These unconventional behaviors echo those of an extensively investigated, but still elusive, series of rare-earth cage compounds : the rare-earth hexaborides. Among them, the most enigmatic case is that of CeB$_6$, which also features a non-magnetic ordering \cite{Effantin1985, Amara2012}.

\begin{figure}
\includegraphics[,width=\columnwidth]{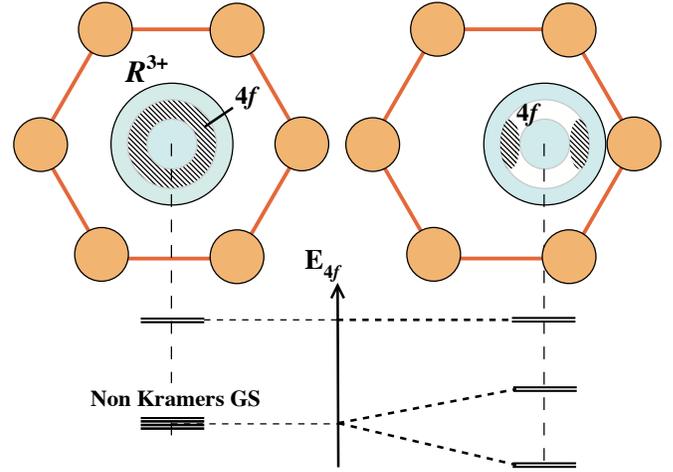}
\caption{\label{CagDiag} 2D schematic of the cage Jahn-Teller mechanism. As the rare-earth ion (R$^{3+}$) goes from the center (left), to an off-center position (right), a non-Kramers level is split (energy scale E$_{4f}$), with a simultaneous angular redistribution of the 4$f$ electrons (hatched area).}
\end{figure}

To understand the properties of rare-earth compounds, accounting for the effect of the Crystalline Electric Field (CEF) is essential. The CEF reflects the anisotropic environment of the 4$f$ ion and lifts, at least partially, the orbital degeneracy of the 4$f$ shell.  As a result, the 2$J$+1 degenerate ground state multiplet is decomposed to form the CEF scheme, according to the point symmetries of the rare-earth site \cite{bethe1929}. In many instances of rare-earth cage compounds, this mandatory step of the analysis cannot be achieved as these systems also display CEF anomalies, revealed by neutron spectroscopy. The inelastic spectra of PrRu$_4$P$_{12}$ are inconsistent with the $T_h$ symmetry of the Pr site below the MI transition and show a spectacular broadening of the CEF excitations above \cite{Iwasa2005}. In PrFe$_4$P$_{12}$, well defined CEF excitations appear only in the ordered state \cite{Park2008}, while, in PrOs$_4$P$_{12}$, they vanish very rapidly with increasing the temperature \cite{Iwasa2009}. In the paramagnetic range of CeB$_6$, the cubic CEF quadruplet ground state appears to split over an energy range of about 30 K \cite{Zirngiebl1984, Loewenhaupt1985}.\\
These recurrent anomalies force to reconsider the effect of the CEF in the cage context. Indeed, how relevant is an approach based on the point symmetries at the cage center, whereas the magnetic atom can be substantially offset? Moreover, in the here considered systems, the point symmetry at the cage center is high. In such circumstances, the CEF ground state is likely to display an excess of orbital degeneracy with respect to the Kramers' minimum. Such an orbital degeneracy can cause a Jahn-Teller (JT) instability: the system's electrostatic energy can be reduced thanks to a distortion that lifts the degeneracy of its electronic states. Many instances of Jahn-Teller effect are thus found in systems where 3$d$ ions occupy sites of octahedral symmetry, such as in perovskite and spinel structures, in which they cause a distortion of the octahedra and, collectively, of the crystal \cite{VanVleck1939, Opik1957}. This cooperative kind of the Jahn-Teller effect is also found in rare-earth compounds \cite{Gehring1975}, notably in high symmetry insulators. These orbitally driven transitions result from a balance between the 4$f$ electrostatic energy and the elastic energy of the lattice.\\
In crystals where high symmetry cages accommodate loosely bound magnetic ions, the symmetry lowering doesn't require a distortion of the cage and, even less, of the crystal : it can be simply achieved by offsetting the magnetic ion (see Fig. \ref{CagDiag}). The energy cost of the displacement should be counterbalanced by the splitting of the orbitally degenerate ground state. Accordingly, the Jahn-Teller effect can be expected to develop more frequently in cage systems than in more common structures. The conditions of emergence of this specific JT effect are here examined, for an encaged 4$f$ ion in a metallic compound, and its most direct physical consequences confronted with reported anomalies.\\
To describe the cage JT effect, one starts from a 4$f$ ion at the center of a high symmetry cage, where its electronic ground state degeneracy exceeds the Kramers' minimum (Fig. \ref{CagDiag}, left side). This ground state is supposed well separated from the first excited level whose influence will be neglected. Within these assumptions, one can treat the two extreme cases of CEF schemes at the cage center: the strong crystal field, that isolates a non-Kramers ground state, and the weak crystal field that leaves a 4$f$ ion with the approximate 2$J$+1 degeneracy of the free ion. As the ion moves out of the center, the symmetry of its environment is drastically reduced (right side of Fig. \ref{CagDiag}). Within a Born-Oppenheimer approximation, the 4$f$ wave-functions adapt to the local crystal field, with a simultaneous lifting of the ground state degeneracy. To describe this symmetry lowering, one has to correct the crystal field hamiltonian by a term that develops simultaneously with the distance $r$ to the cage center. The lowest order correction is second, both in the coordinates of the ion and in those, relative, of the 4$f$ electrons: it describes a coupling between the environment and the 4$f$ quadrupoles.
In case of a cubic symmetry at the center, this imposes a unique form for the crystal field correction $\mathcal{H}_{JT}$, here written using the quadrupolar cubic irreducible representations \cite{MorinSchmitt1990}:\\
\begin{widetext}
\begin{equation}
\label{HJT}
\mathcal{H}_{JT}(\bm{r})=-D^{\gamma} [(3 z^2-r^2) O_{2}^{0} + 3 (x^2-y^2) O_{2}^{2}]
-D^{\varepsilon} [(x \cdot y) P_{xy} + (y \cdot z) P_{yz} + (z \cdot x) P_{zx}]
\end{equation}
\end{widetext}
where $x$, $y$ and $z$ are the components, along the cubic axes, of the displacement $\bm{r}$ of the rare-earth ion.  $\{O_{2}^{0}$, $O_{2}^{2}\}$ and $\{P_{xy}$, $P_{yz}$, $P_{zx}\}$ are the quadrupolar operators transforming, respectively, as the $\gamma$ and $\varepsilon$ cubic representations. In the $J$ manifold of the 4$f$ ion, they are conveniently written in terms of Stevens equivalents \cite{Stevens1952}. $D^{\gamma}$ and $D^{\varepsilon}$ are constants that, within a representation, couple the quadrupoles with the displacement. Eq. (\ref{HJT}) has immediate implications :\\
- an off-center position splits the ground state, forces the emergence of 4$f$ quadrupoles and lowers the electrostatic energy : this is the basis of the Centrifugal Jahn-Teller Effect (CJTE).\\
- in a cage, the CEF ground state has an intrinsic width : to a distributed rare-earth ion corresponds a distributed 4$f$ electrostatic energy.\\
- thermally excited vibrations of the guest, by spreading its distribution, further broaden the ground state.\\
If this broadening is competitive with the CEF splitting at the cage center, well defined CEF excitations cannot be detected. This is consistent with anomalies reported in the CEF investigations of many cage materials \cite{Iwasa2005, Park2008, Iwasa2009, Galera2015}.\\
Independently of any magnetic effect, the ion is confined in the cage, which can be formalized by a potential well $V_0(\bm{r})$ that is essentially temperature independent. Instead of the whole ion, it is convenient to consider only the nucleus, of mass $m$, as the mechanical system trapped in the well. The potential energy can be then assimilated to a static mean-field that, in addition to $V_0(\bm{r})$, includes the interaction of aspherical 4$f$ terms with their environment, according to Eq. (\ref{HJT}). As the central ground state splits over an energy scale in relation with the investigated phenomena, this 4$f$ term depends on the temperature. The total potential thus decomposes into:
\begin{equation}
\label{v(r)}
V(\bm{r}, T)=V_0(\bm{r})+V_{4f}(\bm{r}, T)
\end{equation}
Solving the time independent Schr\"{o}dinger equation at a given temperature requires knowledge of $V(\bm{r})$. This is impractical and one has to resort to approximations.\\
To highlight the CJTE, considering the detail of the cage potential is unnecessary. If the cage ressembles a regular polyhedron, the inside potential created by the point charge vertices is essentially constant, except close to the vertices. Before the RE ion reaches such outer positions, it will be stopped by the strong repulsive forces due to the overlap between its peripheral electrons and those of the cage: the cage framework contribution to the potential amounts to a flat bottom bordered by an almost spherical barrier. In a metallic system, the moving ion also interacts with the valence electrons, which tends to recall it to the cage center. The $V_0(\bm{r})$ potential is thus simplified to a spherical form with two essential features: \\
- an infinite barrier at radius $a$, defining the practical units for length, $a$, and energy, $e.u.=\frac{\hbar^2}{2ma^2}$.\\
- a restoring force, restricted to a quadratic term in the potential for $r < 1$ : $V_0 (\bm{r})= \alpha \; r^2$, where $\alpha$ is positive.\\
The $V_{4f}(\bm{r}, T)$ term essentially reflects the electrostatic interactions between the 4$f$ electrons and the valence electrons. In the present spherical simplification, it should not reflect a quadrupolar anisotropy (see Eq. (\ref{HJT})) and one has to restrict to the case of a negligible CEF at the center, with a $2J+1$ degenerate ground state. Then, taking the $z$ axis along the displacement $r$ direction, the correction of Eq. (\ref{HJT}) is reduced to :
\begin{equation}
\label{HJTr}
\mathcal{H}_{JT}(r)=-  \;D \; r^2 \;O_{2}^{0}=-  \;D \; r^2\;[3J_z^2-J(J+1)] 
\end{equation}
where $D = 2\;D^{\gamma} = 24\;D^{\epsilon}$. \\
To characterize the Jahn-Teller instability, we first focus on the zero temperature limit: then, for a given $\bm{r}$, only the local 4$f$ ground state has to be considered. Depending on the sign of $D$, this ground state will either correspond to the maximum projection doublet, $J_z = \pm J$ or to the minimum, $J_z = 0$ singlet or $J_z =\pm \frac{1}{2}$ doublet. Accordingly, the magnetic entropy is drastically reduced and the magnetic behaviors will differ from those of the 2$J$+1 multiplet. In the process, an $O_{2}^{0}$ quadrupole develops along the displacement axis with eigenvalue $Q$ and the 4$f$ energy change identifies with a centrifugal term: 
\begin{equation}
\label{E4f+}
V_{4f}(r,0)=-  D \; Q \;r^2 = -\beta \;r^2
\end{equation}

\begin{figure}
\includegraphics[,width=\columnwidth]{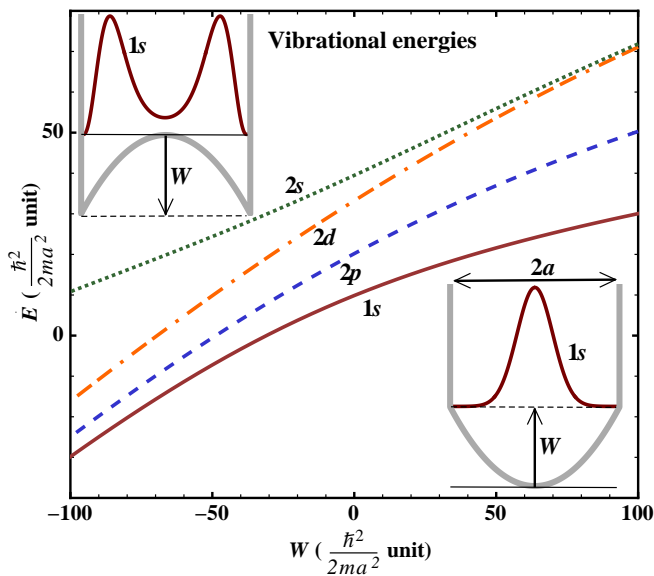}
\caption{\label{VarW} Energies of the 4 lowest oscillator levels as functions of $W$, that defines the quadratic bottom of the potential well. The Jahn-Teller effect shifts $W$ towards negative values. The insets show sketches of the cage potential, and diameter cuts of the 1$s$ distribution for the two extreme values of $W$.}
\end{figure}

\begin{figure}
\includegraphics[,width=\columnwidth]{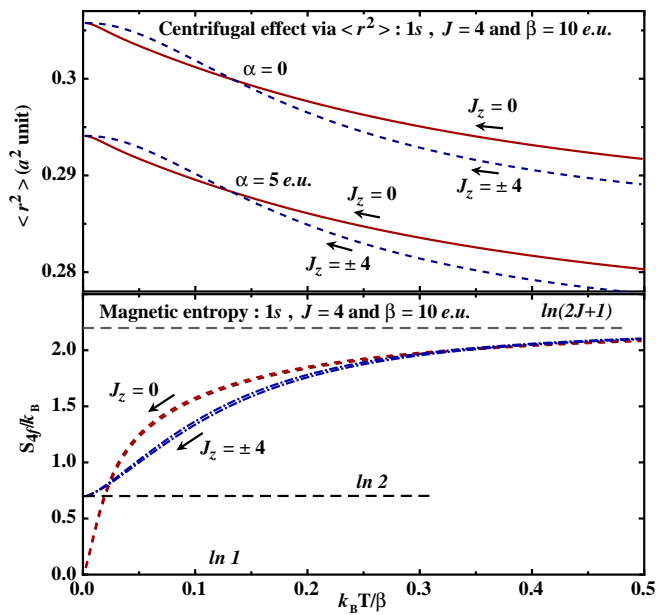}
\caption{\label{VarT} Non-adiabatic treatment of the CJTE spherical model applied to the 1$s$ state, for $J = 4$, $\beta = 10\;e.u.$ and for two values, 0 and 5 $e.u.$, of $\alpha$. For $D <0$ (full line), a local singlet $J_{z} =0$ ground state is favored, while, for $D > 0$ (dashed line), it is the doublet $J_{z} =\pm4$. Upper part : the centrifugal effect through the average of $r^2$. Lower part : the average magnetic entropy as function of the reduced temperature $k_B$T$/\beta$.}
\end{figure}

At $T=0$, for $r<1$, the total cage potential writes as:
\begin{equation}
\label{TotPot}
V(\bm{r}, 0)=(\alpha-\beta )\;r^2=W\; r^2 
\end{equation}
and becomes infinite for $r=1$. $V(\bm{r},0)$ is shown in the insets of Fig. \ref{VarW} for positive and negative $W$. For a given $W$, solving numerically the time independent Schr\"{o}dinger equation yields the wave-functions and energies of the oscillator. Fig. \ref{VarW} shows the dependence on $W$ of the four lowest levels, 1$s$, 2$p$, 2$d$ and 2$s$, in atomic notations. For large $W$ values, the system tends to harmonicity, with a constant level separation. The Jahn-Teller term has an opposite effect, shifting $W$ towards negative values by the amount of $\beta$, which results in : \\
- a reduced energy difference between the three first levels, particularly between the singlet 1$s$ ground state and the first excited triplet 2$p$ states.\\
- a wave-function that progressively adapts to the centrifugal term, the 1$s$ ground state being the most affected (see the insets of Fig. \ref{VarW}).\\
The strength of the centrifugal effect depends on $\beta$, that represents the decrease in the 4$f$ ground state energy for an ion "pressed" against the cage. In such a highly anisotropic environment, the CEF splitting in rare-earth intermetallics can reach a few tens of meV. This energy scale is competitive with the excitation of the first vibration mode observed in cage systems, that ranges from 3.4 meV in PrOs$_4$Sb$_{12}$ \cite{Iwasa2007} to 11 meV in PrB$_6$ \cite{Kohgi2006}. In presence of the CJTE, the cage oscillator properties should be therefore substantially affected. The reduced energy separation between between the 1$s$ and 2$p$ levels means that a softening can be expected for the lowest energy phonons. A possibly related effect is observed in PrOs$_4$P$_{12}$ and PrRu$_4$P$_{12}$ \cite{Iwasa2007}.
The centrifugal effect is mitigated by the restoring force of strength $\alpha$, which essentially depends on the number and distribution of the valence electrons. As a result, the magnitude of the CJTE should vary greatly from one system to another.\\
Another consequence of the CJTE is a higher sensitivity of the ion's distribution to its environment: by reducing the energy spacing and increasing the wave-functions overlap between the lowest levels, the cage multipolar susceptibilities from the $1s$ level are increased. As the oscillator has close levels of different parities, unlike the 4$f$ shell, not only even multipoles, but also odd multipoles can develop in the distribution of the guest ion. This relative freedom in the distribution can be used by the system to minimize its energy when subjected to a magnetic field, mechanical stress, phase transition etc.. For instance, in the antiferromagnetic range of rare-earth hexaborides, cage dipoles (i.e. displacements) play a key role in the definition of the ordered state \cite{Amara2010}. Furthermore, although here neglected, the couplings between the cages exist in real systems. Then, these large multipolar susceptibilities could give rise to non-magnetic orderings of cage multipoles.\\
The above model is defined in the limit of zero temperature: the only considered 4$f$ electronic state is the ground state defined by the local CEF. As the temperature is raised, excited 4$f$ states should interfere in the potential well. The simplest non-adiabatic Born-Oppenheimer approximation consists in assuming that the nucleus is almost motionless from the 4$f$ electrons point of view, allowing them to "explore" the set of locally available CEF states. The massive nucleus filters out the electronic fluctuations and experiences a static mean field consistent with a Boltzmann statistics of the electronic states. As it moves, it exchanges work with the thermalized 4$f$ electrons, which results in a corresponding change in their free energy. In Eq. (\ref{v(r)}), the 4$f$ part of the cage potential, $V_{4f}(\bm{r}, T)$, can be thus identified with the free energy of the 4$f$ electrons. The resulting temperature dependent potential is consistent with the zero temperature limit.  At higher temperatures, all local CEF states get substantially populated, canceling the benefit of an off-center position. Solving the Schr\"{o}dinger equation then yields temperature dependent wave-functions. Fig. \ref{VarT} shows the results of calculations in the example of a $J = 4$ multiplet, at temperatures allowing to neglect the influence of all the vibration levels except the $1s$ ground state. The chosen value $\beta = 10\; e.u. $ is achieved for two values of the coupling constant $D$ (see Eq. (\ref{HJTr})), negative and positive, which respectively favor a singlet, $J_{z} =0$, or doublet, $J_{z} =\pm J$, local ground state. The temperature dependence of $r^2$ averaged over the $1s$ distribution illustrates the centrifugal effect (upper part of Fig. \ref{VarT}). This effect resists an increase in the restoring force constant, from $\alpha=0$ to 5 $e.u$, despite an overall more localized 4$f$ ion. The average magnetic entropy (lower part of Fig. \ref{VarT}), reflects the progressive lifting of the 4$f$ degeneracy and is even less sensitive to $\alpha$. Similar variations of the magnetic entropy within the paramagnetic range, inconsistent with a central CEF scheme, are observed in filled skutterudites \cite{Maple2002} and, spectacularly, in CeB$_6$ \cite{Fujita1980, PEYSSON1986}.

$Summary$: In case of an orbitally degenerate 4$f$ ground state at the center of the cage, the properties of the guest ion vary dramatically with the temperature in the paramagnetic range. Due to the couplings between the 4$f$ ion displacement and its quadrupoles, the degeneracy can be individually and progressively lifted, as the temperature is decreased. Simultaneously, the probability of presence of the magnetic ion increases outside the cage center. This scenario is consistent with anomalies separately observed in the properties of some filled skutterudites and hexaborides compounds. However, to ascertain the occurrence of the CJTE in a given system, one should collect a consistent set of experimental anomalies. In this regard, one can think of the following investigations:\\
- X-ray diffraction, that can probe the distribution of the 4$f$ ion inside the cage. The CJTE, as it spreads this distribution, will result in an anomalous temperature variation of the Debye-Waller factors.\\
- thermal expansion : the radial distribution of the 4$f$ ion impacts the volume of the cage and, consequently, of the crystal.\\
- specific heat : one should observe a progressive reduction in the magnetic entropy while cooling the sample.\\
- magnetic measurements : as it lifts the orbital degeneracy, the CJTE will reduce the effective paramagnetic moment. \\
- spectroscopy of the magnetic excitations : the CJTE can result in an absence of identifiable CEF excitations. In case of a sizable CEF splitting at the center, an energy shift and change in width should be expected.\\
- spectroscopy of the phonons : the vibrations of the guest ions were here treated individually, showing that the CJTE lowers the first excited level. Collectively, this will translate in a softening of the lowest energy phonons.

%

\end{document}